\documentclass[aps,prl,showpacs,superscriptaddress,twocolumn]{revtex4}

\usepackage{hyperref}
\usepackage{bbm}
\usepackage{graphicx}
\usepackage{amsmath}
\usepackage{amssymb}
\usepackage{mathscinet}
\usepackage{amsfonts,amsmath}

\usepackage[matrix,frame,arrow]{xypic}

\begin{document}

\title{
Comments on: Asymptotic Bound for Heat-Bath Algorithmic Cooling }

\author{Nayeli Azucena Rodr\'iÂ­guez-Briones}
\email{narodrig@uwaterloo.ca}
\affiliation{Institute for Quantum Computing, University of Waterloo, Waterloo, Ontario, N2L 3G1, Canada}
\affiliation{Department of Physics \& Astronomy, University of Waterloo, Waterloo, Ontario, N2L 3G1, Canada}
 
 \author{Jun Li}
 \affiliation{Hefei National Laboratory for Physical Sciences at Microscale and Department of Modern Physics, University of Science and Technology of China, Hefei, Anhui 230026, China}


\author{Xinhua Peng}
\affiliation{Hefei National Laboratory for Physical Sciences at Microscale and Department of Modern Physics, University of Science and Technology of China, Hefei, Anhui 230026, China}
\affiliation{Synergetic Innovation Center of Quantum Information \& Quantum Physics,
University of Science and Technology of China, Hefei, Anhui 230026, China}

\author{Tal Mor}
\affiliation{Computer Science Department, Technion, Haifa 320008, Israel}

\author{Yossi Weinstein}
\affiliation{Computer Science Department, Technion, Haifa 320008, Israel}

\author{Raymond Laflamme}
\email{laflamme@iqc.ca}
\affiliation{Institute for Quantum Computing, University of Waterloo, Waterloo, Ontario, N2L 3G1, Canada}
\affiliation{Department of Physics \& Astronomy, University of Waterloo, Waterloo, Ontario, N2L 3G1, Canada}
\affiliation{Perimeter Institute for Theoretical Physics, 31 Caroline Street North, Waterloo, Ontario, N2L 2Y5, Canada}
\affiliation{Canadian Institute for Advanced Research, Toronto, Ontario M5G 1Z8, Canada}

\date{\today}

\maketitle

In a recent paper \cite{Raesi:2014kx}, Raeisi and Mosca gave a limit for cooling with Heat-Bath Algorithmic Cooling (HBAC). Here we show how to exceed that limit by having correlation in the qubits-bath interaction.

Schulman, Mor and Weinstein (SMW)~\cite{schulman2005physical,schulman2007physical} were the first to (wrongly) claim that the Partner Pairing Algorithm (the PPA, which consists of iterations of SORT + free-relaxation) will achieve the optimal physical cooling of HBAC.
SMW provided bounds on the cooling levels that PPA could obtain, but the bounds were not tight, hence it was not possible to observe that PPA is not optimal. SMW assumed that the system-bath interaction proceeds through a swap of qubits, but it turns out that more general form of couplings to the bath exist.

For two qubits (one target qubit, which is going to be cooled, and one reset qubit), starting in the maximally mixed state, the PPA gives a steady state with the qubits at the temperature of the bath and no polarization gain (above the one from the bath) is observed. The first refresh step polarizes the reset qubit, then the first PPA entropy compression transfers that polarization to the target qubit. The reset qubit is then in a fully mixed state and can be repolarised by a thermal contact with the heat bath. It turns out that, after these steps, the diagonal terms are already ordered with non-increasing probability (SORT), such that PPA does not increase the polarization anymore. We thus end up with two qubits in a thermal state, the same as the bath. 

In a recent paper\cite{Li:2014ys}, Jun Li and collaborators studied the efficiency of Liouville space polarization transfer in the presence of a bath, and showed cases/experiments where utilizing relaxation effects does offer and enhancement. In looking at the maximum polarization (or purity), they (re)-discovered that it is possible to enhance the polarization of one of two qubits beyond the bath polarization in presence of relaxation and cross relaxation of the quantum system. This effect was discovered by Overhauser in 1953 \cite{Overhauser:1953fk} and has been observed many times experimentally. It is possible to enhance the polarization of one spin (qubit) at the expense of a second spin (qubit) when their coupling to the bath leads to cross relaxation \cite{Solomon:1955uq}. In the limit of low polarization, the expectation of the Z Pauli operator $S_Z^1$ ($\langle S^1_z \rangle$), obeys the equation (see \cite{Solomon:1955uq}):

\begin{equation}
\frac{d \langle S_z^1\rangle}{dt}= -\rho_1 (\langle S_z^1\rangle - \langle S_z^1\rangle_0)
-\sigma (\langle S_z^2\rangle - \langle S_z^2\rangle_0),
\label{szevol}
\end{equation}
where $\langle S_z^i\rangle_0$ is the expectation of $S_z^i$ at equilibrium when the other spin is not driven (not rotated), $\rho_1$ is the relaxation parameter for the first spin, and $\sigma$ is the cross relaxation parameter.

It is possible to drive (rotate) the second spin so that on the relevant timescale (related to $\rho_2$ and $\sigma$) the expectation of $\langle S_z^2\rangle\approx 0$. Then the steady state of eq.\eqref{szevol} implies that

\begin{equation}
\langle S_z^1\rangle = \langle S_z^1\rangle_0 + \frac{\sigma}{\rho_1} \langle S_z^2\rangle_0.
\end{equation}
Note that $\langle S_z^1\rangle $ corresponds to the polarization of the first qubit, this gives an enhancement compared to what the PPA gives, as long as $\sigma /{\rho_1}$ is positive (which happens in nature). 
One way to understand the process from an algorithmic point of view is to realize that the cross relaxation effectively provides a state relaxation/equilibration (``state reset'') between $|11\rangle $ and $|00\rangle$, without touching the other states, analogous to the qubit reset. This form of reset accompanied by a rotation of the second qubit can however boost the polarization of the first qubit beyond what would be obtained by a qubit reset from the bath as in the PPA.

Thus the PPA, at least for two qubits, gives only a lower bound on maximum polarization achievable for HBAC. It is possible to generalize this idea to enhance the polarization of three qubits beyond the PPA, and the details will be provided in a forthcoming paper.

In conclusion, we have presented a Heat Bath Algorithmic Cooling technique that can have a better polarization enhancement than the one obtained by the PPA. As mentioned in \cite{Briones:2014vn}, the polarization achieved using the PPA should be
interpreted as a lower bound on the maximum amount of polarization that can be achieved. Its importance is due to the simplicity of the PPA when the initial state is totally mixed or in an equilibrium thermal state. From that, it is possible to get analytical results that describe both the steady state and its polarization from which we can determine a variety of properties, e.g. how far it is from polarization of one and explicitly show how much resources are needed \cite{Briones:2014vn}. It will be interesting to see if we can generalize the Overhauser effect and what advantages it can give as we increase the number of qubits.


\bibliographystyle{apsrev4-1}
\bibliography{noe-2}

\end{document}